\documentclass[ reprint,
 amsmath,amssymb,
 aps,
floatfix,
]{revtex4-2}

\usepackage{graphicx}
\usepackage{dcolumn}
\usepackage{bm}

\usepackage{caption}
\usepackage{xcolor}
\usepackage{hyperref}

\begin{document}

\preprint{}

\title{Phase Memory of Orbital Angular Momentum in Multiple Scattering Environment}

\author{Igor Meglinski}
 \email{Correspondence: i.meglinski@aston.ac.uk}
 \affiliation{College of Engineering and Physical Sciences, Aston University, Birmingham, B4 7ET, UK}

\author{Ivan Lopushenko}
\email{ivan.lopushenko@oulu.fi}
 \affiliation{Optoelectronics and Measurement Techniques, University of Oulu, P.O. Box 4500, Oulu, FI-90014, Finland}

\author{Anton Sdobnov}
 \email{anton.sdobnov@oulu.fi}
 \affiliation{Optoelectronics and Measurement Techniques, University of Oulu, P.O. Box 4500, Oulu, FI-90014, Finland}

 \author{Alexander Bykov}
  \email{alexander.bykov@oulu.fi}
 \affiliation{Optoelectronics and Measurement Techniques, University of Oulu, P.O. Box 4500, Oulu, FI-90014, Finland}

\date{\today}

\begin{abstract}

Recent advancements in wavefront shaping techniques have facilitated the study of complex structured light's propagation with orbital angular momentum (OAM) within various media. The introduction of a spiral phase modulation to the Laguerre–Gaussian (LG) beam during its paraxial propagation is facilitated by the negative gradient of the medium's refractive index's temporal change, resulting in an accelerated retardation in OAM twist. This approach attains remarkable sensitivity to even the slightest variations in the medium's refractive index ($\sim 10^{-6}$). The phase memory of OAM is revealed as the ability of twisted light preserving initial helical phase even propagating through the turbid tissue-like multiple scattering medium. The results confirm fascinating opportunities of the exploiting OAM light in biomedical applications, e.g. such as non-invasive trans-cutaneous glucose diagnosis and optical communication through biological tissues and other optically dense media.
\end{abstract}

\maketitle



While using polarised light for the studies of different properties of matter in astrophysics, material science, and biomedicine has already a long history~\cite{Mischenko}, the shaped light possessing Orbital Angular Momentum (OAM)~\cite{Torres} has been added to the potential practical toolkit only recently~\cite{Rubinsztein-Dunlop_2017}. The shaped OAM light plays an emerging role in both classical and quantum science, and offers fascinating opportunities for exploring new fundamental ideas, as well as for being used  in practical applications~\cite{Shen}. The research in the field of shaped light with OAM achieved high recognition in generation and characterization of the exotic vector laser beams~\cite{Rego}, improving telecommunication technologies~\cite{Wang}, optical trapping of cells~\cite{Zhu} and micro-particles~\cite{Padgett2012}. Despite the technical challenges, the shaped OAM light is emerged as one of the most exciting front-lines of contemporary research, promising to defy the deficiencies of current optical sensing techniques~\cite{Weng}. OAM-based twisted light holds great potential for various biomedical applications in the field of biological tissue diagnosis. Apart from the precise optical manipulation and sorting of biological particles and cells utilizing optical optical tweezers~\cite{Bustamante,Avsievich} the emerging biomedical applications based on OAM-based twisted light include imaging and microscopy~\cite{Monika}, generation twisted light from typical on-chip devices~\cite{APL}, optical communication in biological media~\cite{Daryl} and other~\cite{Weng,Chen}.  

We explore the prediction capacities of OAM of Laguerre-Gaussian (LG) beams by analyzing their helical wavefront change along propagation through a tissue-like medium. In frame of the paraxial approximation LG beam is defined as~\cite{Allen1999-uc,Rosales-Guzman2018-fl,Berry2008-ly}:
\begin{equation}
\begin{aligned}
& L G_{p}^{\ell}(\rho, \phi, z)= \sqrt{\frac{2 p !}{\pi(\lvert\ell\rvert+p) ! {w}^{2}(z)}}\left[\frac{\rho \sqrt{2}}{{w}(z)}\right]^{\lvert\ell\rvert} \times\\
& \times L_{p}^{\lvert\ell\rvert}\left[\frac{2 \rho^{2}}{{w}^{2}(z)}\right] \exp \left[-\frac{\rho^{2}}{{w}^{2}(z)}\right] \exp [i(2 p+\lvert\ell\rvert+1) \times\\
& \times\arctan(z/z_R)]  \exp \left[\frac{-i k \rho^{2}z}{2(z^2+z_R^2)}\right] \exp [-i \ell \varphi] \exp [-i kz].
\end{aligned}
\label{eq:LGbeamField}
\end{equation}
Here, $k=2\pi/\lambda$, $\lambda$ is the wavelength of laser radiation, ${w}(z)={w}(0) \sqrt{1+(z/z_R)^2}$, $z_R=\pi {w}^2(0)/\lambda$, ${w}(0)$ corresponds to the zero-order Gaussian beam waist and is adjusted to fit experimental image, $\{ \rho,\phi,z\}$ represents the cylindrical coordinates system utilized for characterization of beam propagation along $z$ axis in terms of radial ($\rho$) and angular ($\phi$) coordinates.  

Phase evolution of the LG beam along its propagation in a medium is defined as~\cite{Berry2008-ly}:
\begin{equation}
\begin{aligned}
\label{eq:LGbeamPhase}
   & \Psi(\rho, \phi, z) = \arg\left(L G_{p}^{\ell}(\rho, \phi, z)\right) = \\
& = \frac{-k \rho^2 z}{2(z^2+z_R^2)} - \ell\phi - kz + G(z), 
\end{aligned}
\end{equation}
where $G(z)=(2p+\lvert\ell\rvert+1)\arctan(z/z_R)$ corresponds to the Gouy phase~\cite{Andrews}. 
The helical phase of an LG beam refers to the phase front that winds around the beam's axis, is a result of the azimuthal phase term in the beam's electric field expression, which corresponds to the topological charge $(\ell)$ and radial index $(p)$.

\begin{figure*}[!t]
\includegraphics[width=6in]{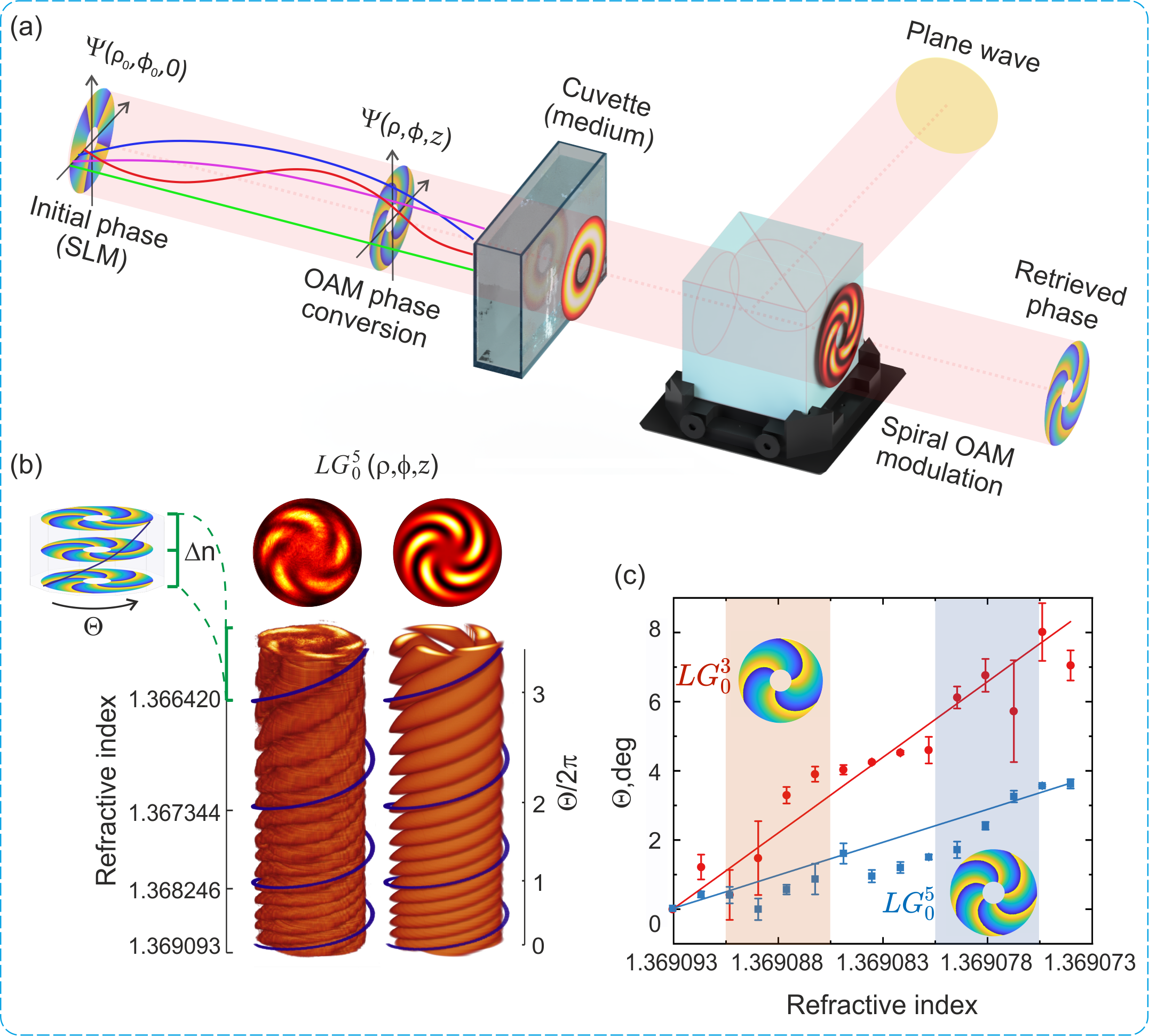}\\
\caption{
  (a) Schematic representation of the central part of the Mach-Zehnder interferometer experiment: the LG beam, carrying OAM imparted by a spatial light modulator (SLM), traverses a cuvette containing the probed medium; interference between the transmitted LG beam and a reference plane wave (an initially expanded Gaussian beam) is then analyzed for intensity and/or retrieved phase distributions.
  (b) Spiral modulation of the intensity of $LG^5_0$ beam observed experimentally (left) during the gradual increase of the medium's refractive index, alongside the prediction from theoretical modeling~\cite{Ivan,Doronin2019} (right); Blue line corresponds to accelerated retardation of OAM twist along LG beam propagation in the medium.
    The inset (left) depicts relative phase twist ($\Theta = \dfrac{\Psi}{\ell}$) of OAM of $LG^5_0$ beam during gradual increase of medium refractive index ($\Delta n=3.69 \times 10^{-4}$). 
    (c= Relative phase twist of OAM of $LG^3_0$ (circles) and $LG^5_0$ (squares) beams during the gradual increase of medium refractive index within a range of $2 \times 10^{-5}$; Highlighted colored areas ($\Delta n = 5 \times 10^{-6}$) feature corresponding twist of OAM for $LG^3_0$ and $LG^5_0$ beams, respectively.}
      \label{fig1}
\end{figure*}

The LG beam retains its helical structure (\ref{eq:LGbeamPhase}) while propagating through the transparent medium, preserving the phase composition $\Psi(\rho,\phi,z)$ associated with OAM and the characteristic helix-like wavefront (Fig.~\ref{fig1}). Here and below we consider, so-called, scalar LG beams~\cite{Forbes_Rev} with homogeneous annular polarisation distribution ($LG^3_0$ and $LG^5_0$).
The observed an accelerated retardation in the OAM rotation along with relative twist of phase ($\Theta = \dfrac{\Psi}{\ell}$) of the LG beam (see Fig.~\ref{fig1}-b), as it propagates through the medium with the negative gradient of the temporal change of the refractive index, are due to spatial dispersion of Poynting vector trajectories, defined as~\citep{Berry2008-ly,Bliokh2013}: 

\begin{equation}
\begin{aligned}
&L(r_0,\varphi_0,\zeta_s) = \int_0^{\zeta_s} \sqrt{ \left(\frac{dr}{d\zeta}\right)^2 + r^2\left(\frac{d\varphi}{d\zeta}\right)^2 } d\zeta, \\
&r(\zeta) = r_0\sqrt{1+4\zeta^2}, \quad \varphi(\zeta)=\frac{\ell}{2 r_0^2}\arctan(2\zeta) + \varphi_0.
\end{aligned}
\label{eq:LGtrj}
\end{equation}

\newpage

\begin{figure*}[!t]
         \includegraphics[width=6in]{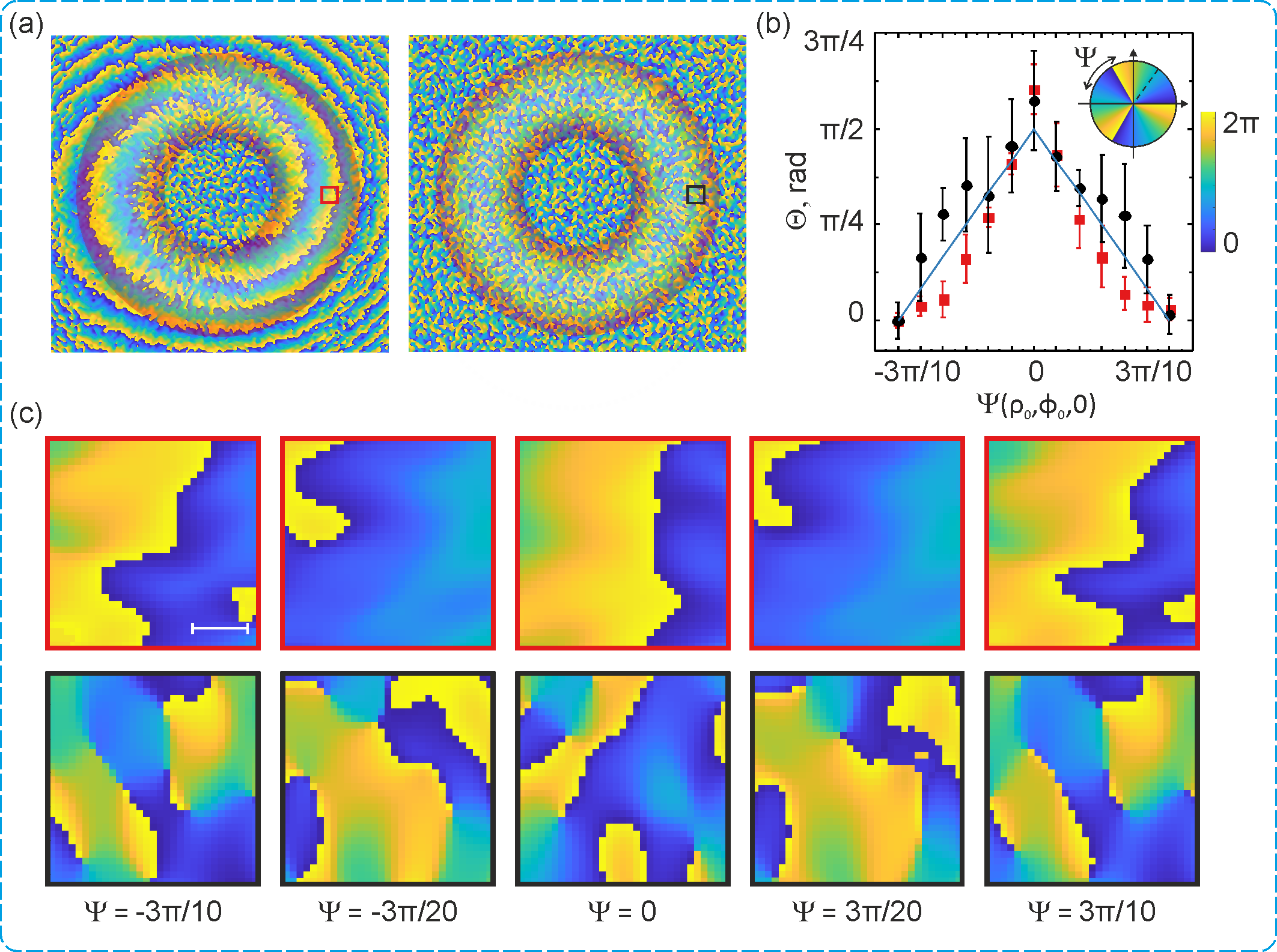}\\
        \caption{ (a) Phase distribution (speckle patterns) observed experimentally for the $LG^3_0$ beam propagated through the low $(d/l^*=2)$ scattering (left) and multiple $(d/l^*=9.6)$ scattering (right) media. The axial annular zone (embossed by contours) corresponds to the $LG^3_0$ beam as if it were passing through a medium devoid of scattering.
   (b) Phase variations manifest at the single speckle grain within a deliberately chosen sector of the $LG^3_0$ axial annular area for low-scattering (indicated by black circles) and multiple-scattering (represented by red squares) media. The observed phase changes are contingent upon the initial phase configuration ($-3\pi/10 \leq \Psi \leq 3\pi/10$) established at SLM (schematically shown in inset). 
   (c) The ensuing alterations in the phase mapping of the speckle pattern within the designated areas ($150 \times 150~\mu m$) highlighted in the $LG^3_0$ axial annular domain (see Fig.~\ref{fig:2}-a), aligning with the prescribed initial phase configuration established at the SLM ($-3\pi/10 \leq \Psi \leq 3\pi/10$). The upper and lower rows present, respectively, scenarios for low and multiple scattering  environment; scale bar corresponds to $150~\mu m$. \href{https://doi.org/10.5281/zenodo.10377315}{Video 1} presents the speckle pattern phase dynamics corresponding to the phase twist at the SLM.   
   }
   \label{fig:2}
\end{figure*}

Here, $L(r_0,\varphi_0,\zeta_s)$ is the length of Poynting vector trajectory along LG beam propagation;
$r_0, \varphi_0$ define the radial and angular coordinates of the trajectory at the starting point ($z=0$); $r=\dfrac{\rho}{{w}(0)}$ and ${\zeta=\dfrac{z}{k{w}^2(0)}}$ are the dimensionless cylindrical coordinates; $\zeta_s$ defines the point to which trajectory length is evaluated; $r(\zeta)$, $\varphi(\zeta)$ are the corresponding coordinates at the trajectory ($0\leq \zeta\leq \zeta_s$), the direction of which ($\mathbf{p}=\{p_\rho, p_\phi, p_z\}$) defined as~\cite{Allen1999-uc,Berry2008-ly}:
\begin{equation}
\begin{aligned}
& p_{\rho}= \frac{\omega k \rho z}{\left(z_{R}^{2}+z^{2}\right)} \lvert LG^\ell_p(\rho,\phi,z)
\rvert^{2}, \\
& p_{\phi}=\frac{\omega l}{\rho} \lvert LG^\ell_p(\rho,\phi,z)\rvert^{2}, \\
& p_{z}= \omega k \lvert LG^\ell_p(\rho,\phi,z)\rvert^{2} \text {, }
\end{aligned}
    \label{eq:LGbeamDirections}
\end{equation}
where $\omega$ is the angular frequency of the light.

In LG beam the Poynting vector trajectories follow the spiral paths around the optical axis of the beam~\cite{Berry2008-ly,Bliokh2013}, as presented in Fig.~\ref{fig1}-a. Gradual increase of the medium's refractive index gives a proportional elongation of these spiral-like trajectories and increased length deviation between them. The increment of the phase component $\left(\frac{-k \rho^2 z}{2(z^2+z_R^2)} - kz \right)$ along the longer trajectories leads to the emergence of a gradient in the LG beam phase (\ref{eq:LGbeamPhase}) that in turn manifests itself as a twisting effect in its transverse distribution (see Fig.~\ref{fig1}-b). Accordingly, when the characteristic size of the medium (e.g. the thickness of the medium) is much larger than the wavelength ($d \gg \lambda$) even a minor variation in the medium refractive index leads to a significant alteration of the transverse spatial phase distribution of the LG beam (see Fig.~\ref{fig1}-c). A prominent twist of OAM is observed experimentally up to minuscule changes of the medium refractive index ($\Delta n = 10^{-6}$ (see Fig.~\ref{fig1}-c).

A gradual decrease in the medium's refractive index induces a spiral modulation of the LG beam, observed via interference with a plane wave (an expanded Gaussian beam), as an accelerated retardation of the OAM twist (see Fig.~\ref{fig1}-b). The experimental results obtained align closely with the theoretical predictions as presented in Fig.~\ref{fig1}-b. In case of the LG beams with lower topological charge ($\ell = 3$), the Poynting vector trajectories tend to less tightly wound around the optical axis~\cite{Berry2008-ly,Bliokh2013}. Therefore, the OAM twist for $LG^3_0$ beams with lower topological charge exhibits higher accuracy in predicting even negligible changes in the refractive index compared to the $LG^5_0$ beam (see Fig.~\ref{fig1}-c).  

While the twist of OAM of LG beam along propagation through a transparent medium (see Fig.~\ref{fig1}) primarily arises due to gradual increase of the medium refractive index, causing a corresponding enhancement of the lengths of spiral the Poynting vector trajectories (\ref{eq:LGtrj}), in turbid medium, the scattering of the LG beam leads to the speckle pattern formation. Arising from the superposition of partial components of the helical wavefront, the speckle interference pattern manifests spatially varying intensity and phase distributions. This occurrence contributes to the disruption of the LG beam helical structure composition in low-scattering media and its complete degradation in multiple scattering media (as presented in Fig.\ref{sup:A}, see Appendix A).

Disperse turbid media exhibiting low or multiple scattering of light are characterised by their optical depth ($\sim d/l^{*}$), which serves as a quantitative measure of the extent of attenuation and scattering strength experienced by light along its propagation through. Here, $d$ is the thickness of the scattering medium and $l^{*}$ is the transport mean free path. A commonly used guideline is that if $d/l^{*} \sim 10$ or larger the medium is considered as multiple or diffuse scattering~\cite{Alfano1990}, whereas for the single and an intermediate (`snake-like photons') scattering $d/l^{*}$ is lower ($\sim 3-6$). 

In a low scattering medium ($d/l^{*} \sim 2$), the LG beam propagates with minimal disruption, enabling it to maintain its initial OAM state and doughnut-like spatial intensity profile~\cite{Doronin2019,Vasilis}, as well as the helical phase structure (Fig.~\ref{fig:2}-a). The multiple scattering ($d/l^{*} \sim 10$) results in a diffusive spread of the LG beam's intensity profile and the destruction of the helical phase front, leading to creation of complex speckle pattern (see Fig.~\ref{fig:2}-a). 
Nevertheless, despite the influence of strong diffuse scattering, the phase speckle pattern maintains the modulation of the initial phase of the LG beam, representing a manifestation of the OAM's memory in multiple scattering. 

The preservation of the helical phase structure by LG beam along propagation through the scattering medium (see Fig.~\ref{fig:2}-b and Fig.~\ref{fig:2}-c) is attributed due to the similarities in the deviations of the Poynting vector trajectories around the beam axis causing due to rotational symmetry. 

\begin{figure*}[!htb]
          \includegraphics[width=6in]{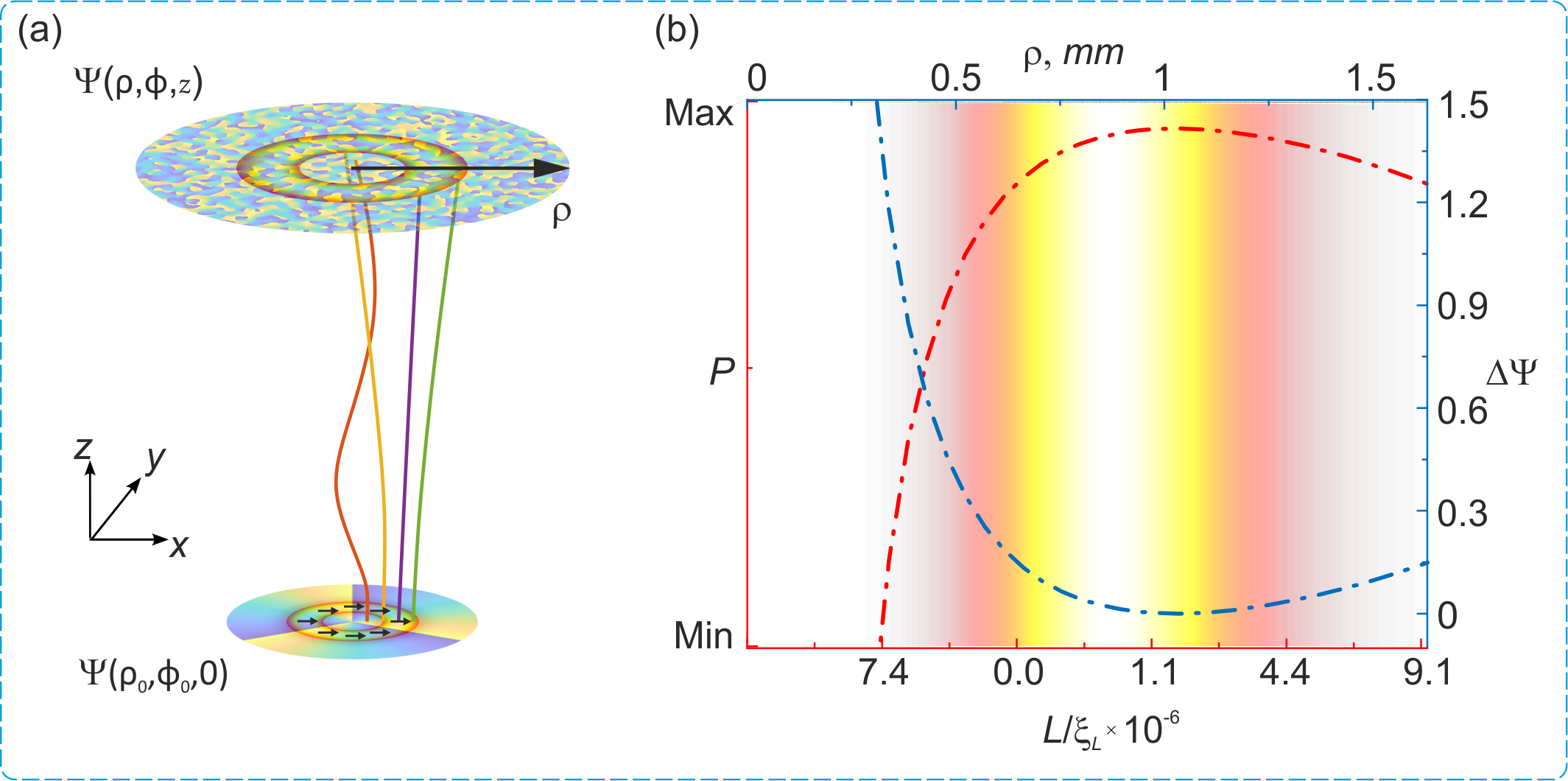}\\
     \caption{(a) The Poynting vector trajectories, schematically depicted in Cartesian coordinates, for the scalar linearly $x$-polarized $LG^3_0$ beam as if it were passing from the SLM to the axial annular zone (embossed by contours) at the detector through a medium devoid of scattering. 
    (b) Radial distribution of the degree of polarization ($P$) and relative phase shift ($\Delta\Psi$) occurring along the open channel as propagated through a multiple scattering medium ($d/l^* \sim 10$) with the characteristic length of depolarization ($\xi_L$). The colourful background represents longitudinal profile of the intensity of $LG^3_0$ beam along the radial component ($\rho$). 
    } 
      \label{fig_DOLP1}
\end{figure*}

Influenced by the scattering medium, the Poynting vector trajectories significantly transform in relation to their original spiral paths (Fig.~\ref{fig_DOLP1}-a), resulting in different partial components of the helical wavefront of the LG beam experiencing varying degrees of phase distortions.

Due to the rotational symmetry, the deviations in the Poynting vector trajectories tend to be similar in magnitudes and directions around the beam axis. 
Therefore, despite the substantial distortions of phase induced by multiple scattering, the overall preservation of OAM of the LG beam is clearly observed in the phase dynamics of the speckle pattern, corresponding to the modulation of the initial phase at the SLM (see Fig.~\ref{fig:2}-c and \href{https://doi.org/10.5281/zenodo.10377315}{Video 1}).  
Remarkably, during propagation through the multiple scattering medium ($d/l^{*} \sim 10$), the phase memory of OAM is distinctly pronounced within the axial annular region of the LG beam, as embossed in Fig.~\ref{fig:2}-a (see also \href{https://doi.org/10.5281/zenodo.10377315}{Video 1}), with no discernible presence in the central and external areas beyond the annular region. 

When light passes through a turbid tissue-like medium, even with a low scattering ($d/l^{*} < 2$), the wavefront undergoes rapid deformation, resulting in a distinct speckle pattern. Notably, while traversing a more scattering medium, polarization scrambling occurs on a different length scale~\cite{Brasselet}, leading to the adoption of a chaotic or disordered nature in the polarization state of light.  

In other words, the initial orientation of the electric field vector representing the polarization of light undergoes unpredictable changes due to interactions with scattering elements in the medium. Polarization scrambling results in a loss of the initial alignment, leading to a more random distribution of polarization states. The degree of polarization of light is defined as the measure of the alignment of the electric field vectors within the light wave~\cite{Bicout}:
\begin{equation}
\begin{aligned}
P=\dfrac{2L}{l_{s}}\mathrm{sinh}\left(\dfrac{l_{s}}{\xi_L}\right)\exp\left(-\dfrac{L}{\xi_L}\right) \text {, }
\end{aligned}
    \label{eq:DoP}
\end{equation}
where $\xi_L$ is the characteristic depolarization length $\left(\xi_L=l_{s}/\left(\sqrt{3\mathrm{ln}\frac{10}{7}}\right)\right)$, $l_s$ is the elastic mean free path ($l_s=l^{*}(1-\langle\cos\theta\rangle)$), $\theta$ is the scattering angle, and $L$ corresponds to the Poynting vector trajectories (\ref{eq:LGtrj}) of LG beam (see Fig.\ref{fig_DOLP1}-a). 

In disordered scattering media, such as biological tissues and turbid tissue-like materials, light encounters multiple scattering events with particles or structures~\cite{Jacques_2013}, leading to the formation of branched transport channels~\cite{PNAS}. In a similar fashion, the preservation mechanism within the axial annular region is associated with the exponential localization of the transmission eigenchannels, aligning with the shortest optical pathway.~\cite{Hasan}.

Fig.~\ref{fig_DOLP1}-b shows gradual transformation of polarization degree along the radial component ($\rho$) of the $LG^3_0$ beam, corresponding to the spread of the pathways ($L/\xi_L$) within the medium from SLM to the detecting area as presented in Fig.~\ref{fig_DOLP1}-a. In other words, upon propagation of the scalar linearly $x$-polarized $LG^3_0$ beam through multiple scattering medium ($d/l^* \sim 10$) the polarization remains effectively along the eigenchannel, whereas the relative phase distortion $\Delta\Psi=\dfrac{\Psi\vert_{L/\xi_L}-\Psi\vert_{z/\xi_L}}{2\pi}$ becomes minimal within the axial annular zone (see  Fig.~\ref{fig_DOLP1}-b).  

Eventually, upon propagation through a turbid tissue-like scattering medium ($d/l^* \sim 10$), the phase memory of Orbital Angular Momentum (OAM) exhibits spatial variability when examined along $\rho$ — the radial position (schematically shown in Fig.~\ref{fig_DOLP1}-a). The overall helical phase structure associated with the OAM remains preserved, clearly observed in the single speckle grains within an arbitrarily chosen area within the annular region of the LG beam (see Fig.~\ref{fig:2}-a). However, in the areas outside of the doughnut-like ring, the phase memory effect disappears. Conversely, in a low-scattering environment ($d/l^* \sim 2$), the phase memory is more consistently observed in all regions (see Fig.\ref{sup:B} in Appensix B, and \href{https://doi.org/10.5281/zenodo.10377389}{Video 2}).

Computational analyses corroborate these experimental findings, affirming that the phase memory of OAM in LG beams propagated through turbid tissue-like scattering media is not absolute. It is susceptible to various factors, including scattering, absorption, and anisotropy of scattering, potentially leading to the degradation or alteration of the OAM's phase content. 

The obtained results offer fascinating opportunities of the exploiting OAM light in biomedical applications, e.g. such as non-invasive trans-cutaneous glucose diagnosis and optical communication through biological tissues and other disperse multiple scattering materials.

\subsection*{Optical setup and processing}

The experimental configuration employed for measurements is illustrated in Fig\,\ref{fig4} in Appendix C. In summary, a Gaussian light beam at $640~nm$ is divided into sample and reference beams using a polarizing beam splitter. A spatial light modulator (PLUTO-2-NIR-011, Holoeye, Germany) is utilized in the reference arm to generate various LG beams carrying OAM. These OAM-carrying beams traverse the sample -- scattering environment (refer to Appendix D for details). Simultaneously, the reference Gaussian beam undergoes expansion through a series of lenses and is directed to a beam splitter, where it interferes with the sample beam. Subsequently, the interference pattern is recorded using a charge-coupled device (CCD) camera. To establish a comparison between experimental data and theoretical simulations, the methodology outlined in Appendix E is employed.

The utilized setup facilitates the acquisition of interference patterns in both on-axis~\cite{cui2019determining} and off-axis~\cite{vayalamkuzhi2021transform} regimes. The determination of relative OAM twist in the on-axis regime involves assessing the change in polar angle of a petal caused by its rotation around the LG beam center. Conversely, in the off-axis regime, the LG beam phase is retrieved through a fast Fourier transform approach~\cite{vayalamkuzhi2021transform}. A detailed elucidation of the experimental data processing methodology is presented in Appendix F and G, respectively.

\begin{acknowledgments}
Authors acknowledge the support from the Leverhulme Trust and The Royal Society (Ref. no.: APX111232 APEX Awards 2021), UKKi UK-Israel innovation researcher mobility and Academy of Finland (grant projects 325097, 351068).
\end{acknowledgments}

\bibliography{main}

\begin{figure*}[!htbp]

\textbf{APPENDIX A: Addition to Figure 2}\vspace*{5mm}
 \includegraphics[width=6in]{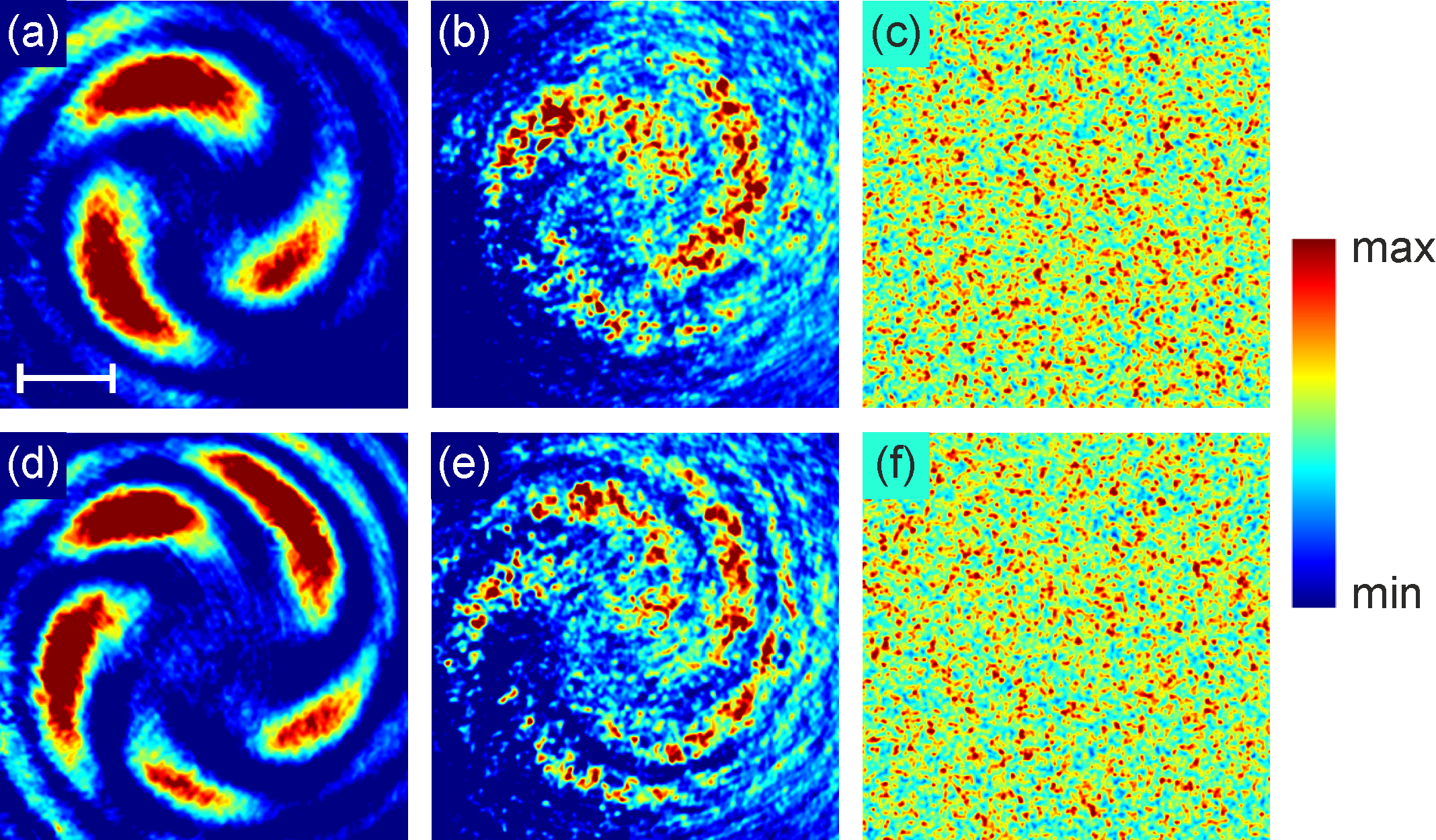}\\
  \renewcommand{\thefigure}{A1}
  \centering 
   \caption{ Experimentally observed spatial intensity distributions for $LG^3_0$ and $LG^5_0$ beams propagated through a transparent medium (a) and (d), as well as through the low $(z/l^*=2)$ and multiple $(z/l^*=9.6)$ scattering environments, respectively, (b) and (e) and (c) and (f). Scale bar is equal to $750~\mu m$.} 
      \label{sup:A}
\end{figure*}

\begin{figure*}[!ht]
\renewcommand{\thefigure}{B1}
 \centering
 \textbf{APPENDIX B: Addition to Video 2}\vspace*{5mm}
 \includegraphics[width=6in]{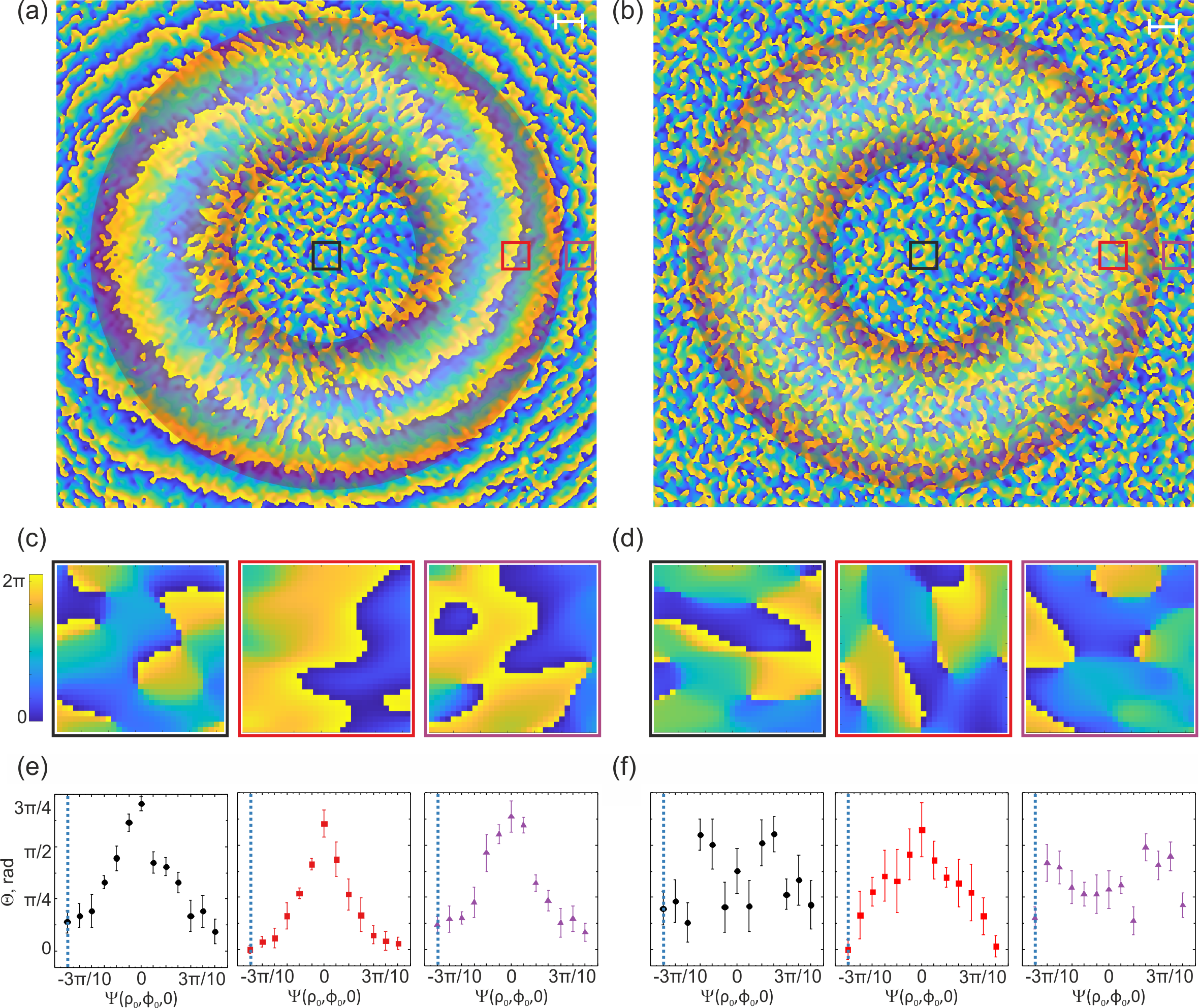}\\
 \renewcommand{\thefigure}{B1}
   \caption{ Phase distribution (speckle patterns) observed experimentally for the $LG^3_0$ beam propagated through the low $(d/l^*=2)$ scattering (a) and multiple $(d/l^*=9.6)$ scattering (b) media. The axial annular zone (embossed by contours) corresponds to the $LG^3_0$ beam as if it were passing through a medium devoid of scattering. 
    The resulting changes in the phase distribution of speckle patterns observed within specified areas (size: $150 \times 150~\mu m$) selected at the center (left), axial annular region (center), and outside (right) of the $LG^3_0$ beam profile (as indicated in (a) and (b), presented for both low scattering (c) and multiple scattering (d) environments.
    Respectively, the phase variations observed at the single speckle grain within  center (left), axial annular region (center), and outside (right) of the $LG^3_0$ beam both low (e) and multiple (f) scattering. The observed phase changes are contingent upon the initial phase configuration ($-3\pi/10 \leq \Psi \leq 3\pi/10$) established at SLM see  \href{https://doi.org/10.5281/zenodo.10377389}{Video 2}.} 
      \label{sup:B}
\end{figure*}

\clearpage

\subsection*{APPENDIX C: Experimental setup}
\begin{widetext}

 \begin{figure*}[!ht]
 \centering
 \renewcommand{\thefigure}{C1}
 \includegraphics[width=6in] {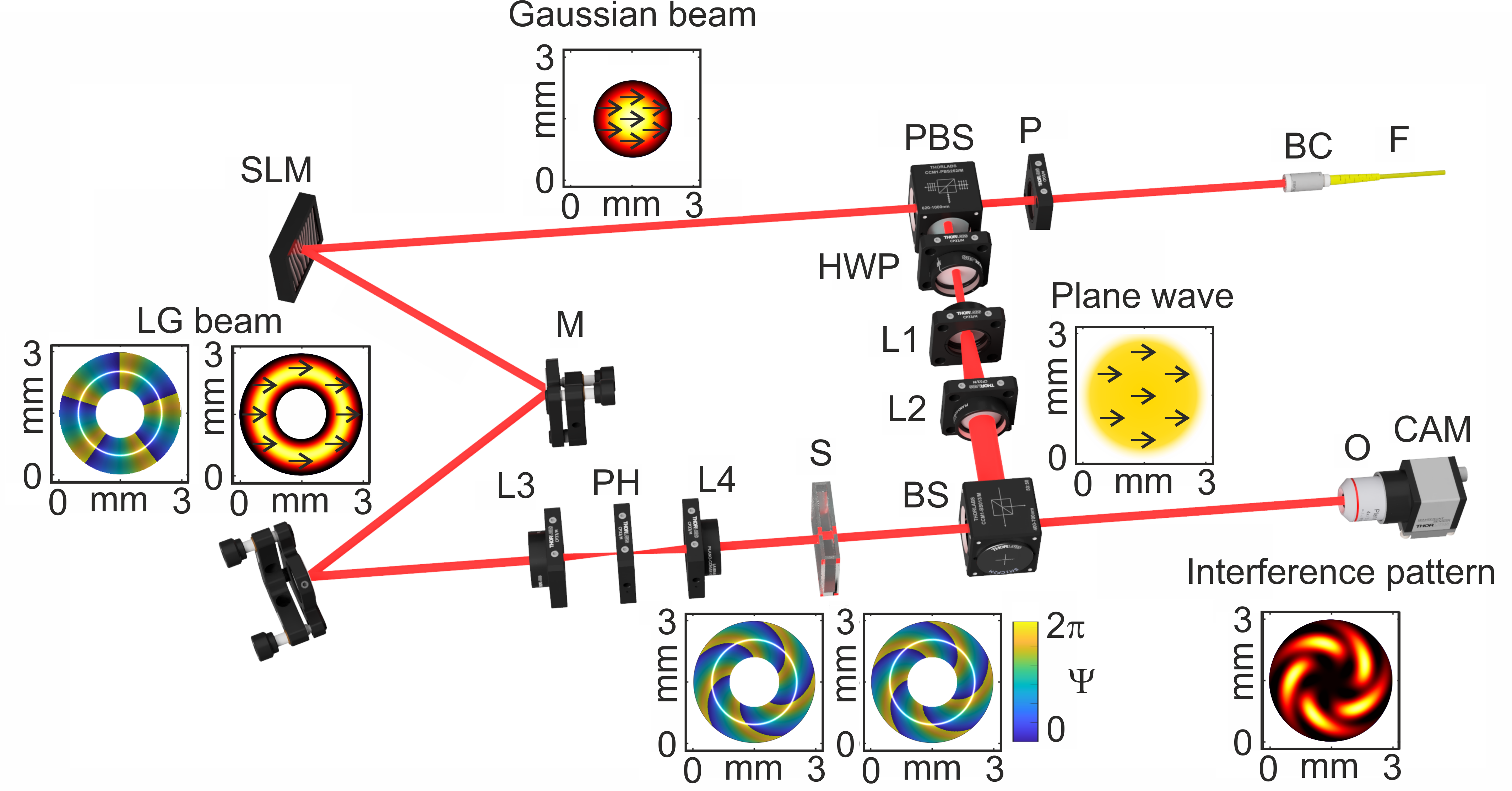}\\
   \caption{LD -- laser diode. P -- polarizer. FM -- fiber mount. BC -- beam colimator. SLM -- spatial light modulator. M1, M2 -- mirrors. L1, L2, L3, L4 -- lenses. PH -- pinhole. PBS -- polarizing beam splitter. HWP -- half wave plate. S -- cuvette filled with the sample liquid. BS -- beam splitter. NF -- neutral filter. O -- objective. CCD -- camera. The detailed description of the optical setup is presented in the main article text.} 
      \label{fig4}
\end{figure*}
\end{widetext}

The modified Mach–Zehnder-based interferometer~\cite{kumar2019modified} is used to examine an evolution of OAM of the LG beams propagated through the multiple scattering environment (Fig.~\ref{fig4}).

A coherent Gaussian light beam ($40~mW$, BioRay laser diode, Coherent, USA with coherence length $>20~cm$) emitting at $640~nm$ serves as the light source. To clear the optical mode, the laser beam is focused into a single-mode optical fiber (Thorlabs, USA) F. The resulting output beam is collimated using a beam collimator (Thorlabs, USA) BC to obtain a Gaussian beam with a $1.6~mm$ waist diameter. Additionally, a polarizer (Thorlabs, USA) P is used after the collimator to achieve horizontal linear polarization. The Gaussian beam is then split into sample and reference beams using a polarizing beam splitter (Thorlabs, USA) PBS.

The sample beam illuminates the phase-only spatial light modulator (PLUTO-2-NIR-011, Holoeye, Germany) SLM, operating in reflective regime. To produce LG beams with different moments, the corresponding forked diffraction patterns are generated on the SLM. The diffracted light from the SLM is directed using a set of mirrors M to lens L3 ($f=45~mm$, Thorlabs, USA). This lens is used to focus the first-order diffraction through a pinhole (Thorlabs, USA). The LG beam is then re-collimated using lens L4 ($f=45~mm$, Thorlabs, USA). Finally, the sample beam passes through the sample S.
The reference beam goes through a half-wave plate (Thorlabs, USA) to control the polarization orientation of the Gaussian beam. This reference beam is expanded by lenses L1 ($f=30~mm$ Thorlabs, USA) and L2 ($f=70~mm$, Thorlabs, USA), and directed to the beam splitter (Thorlabs, USA) BS, where the expanded Gaussian beam interferes with the LG beam. The interference pattern is then registered using a CMOS camera (DCC3240M, $1280 \times 1024$, Thorlabs, USA) CAM in combination with an objective (10$\times$, Nikon, Japan) O. The waist beam diameter for $LG^3_0$ and $LG^5_0$ beams at the detector is $2.7$ and $3~mm$, respectively. 

The setup enables the acquisition of interference patterns in both on-axis~\cite{cui2019determining} and off-axis~\cite{vayalamkuzhi2021transform} regimes. In the on-axis scenario, the definition of relative OAM twist involves tracking the change in the polar angle of a petal caused by its rotation around the center of the LG beam. Meanwhile, in the off-axis regime, the retrieval of the LG beam phase is accomplished through a fast Fourier transform approach~\cite{vayalamkuzhi2021transform}.

\subsection*{APPENDIX D: Samples: scattering environment} 
The selection of the ethanol-water solution as an optically transparent medium for experimental inquiry is underpinned by careful consideration. The quantification of the relative OAM twist, stemming from temperature-dependent changes in the refractive index of the solution, is systematically conducted. 
A precisely measured volume of $5~ml$ of an ethanol-water solution, characterized by a water concentration of $51.89~mol\%$, is carefully introduced into a glass cuvette. The cuvette, with a total thickness of $5.6~mm$, features glass walls, each having a thickness of $1~mm$.

The investigation thoroughly addresses the intricate thermal dependencies governing the refractive index of the ethanol-water amalgamation, as comprehensively detailed in the authoritative work~\cite{jimenez2009concentration}. Subsequently, the cuvette containing the liquid sample undergoes refrigeration until reaching a temperature of $7.8 \pm 0.5^\circ C$. After the cooling phase, the cuvette is reintegrated into the experimental apparatus and subjected to a $25~minute$ duration at room temperature ($21^\circ C$). This deliberate protocol induces modulations in the refractive index of the sample liquid during the heating process.

The nuanced alterations in interference patterns, reflective of variations in the refractive index, are meticulously captured by a high-speed camera throughout the entirety of the 25-minute heating interval. The recording is conducted at a frame rate of 10 frames per second, each frame having a $1~msec$ exposure time. These experimental procedures adhere to rigorous standards to ensure the fidelity and reliability of the acquired data.

Throughout the experimental duration, the sample temperature is vigilantly monitored at 10-second intervals utilizing a FLIR C5 thermal camera from Teledyne FLIR, USA.

The measurements are conducted in a low scattering ($d/l^* = 2$) phantom with a thickness of $1~mm$ ($\mu_s=10~mm^{-1}$) and the multiple-scattering one ($d/l^* \sim 10$) with a thickness of $8~mm$ ($\mu_s=6~mm^{-1}$). The meticulous methodology employed in the preparation of the phantoms is comprehensively expounded upon in the work by Wrobel et al., as documented in~\cite{wrobel2015measurements}.

\begin{figure*}[!ht]
\renewcommand{\thefigure}{F1}
     \includegraphics[width=6in]{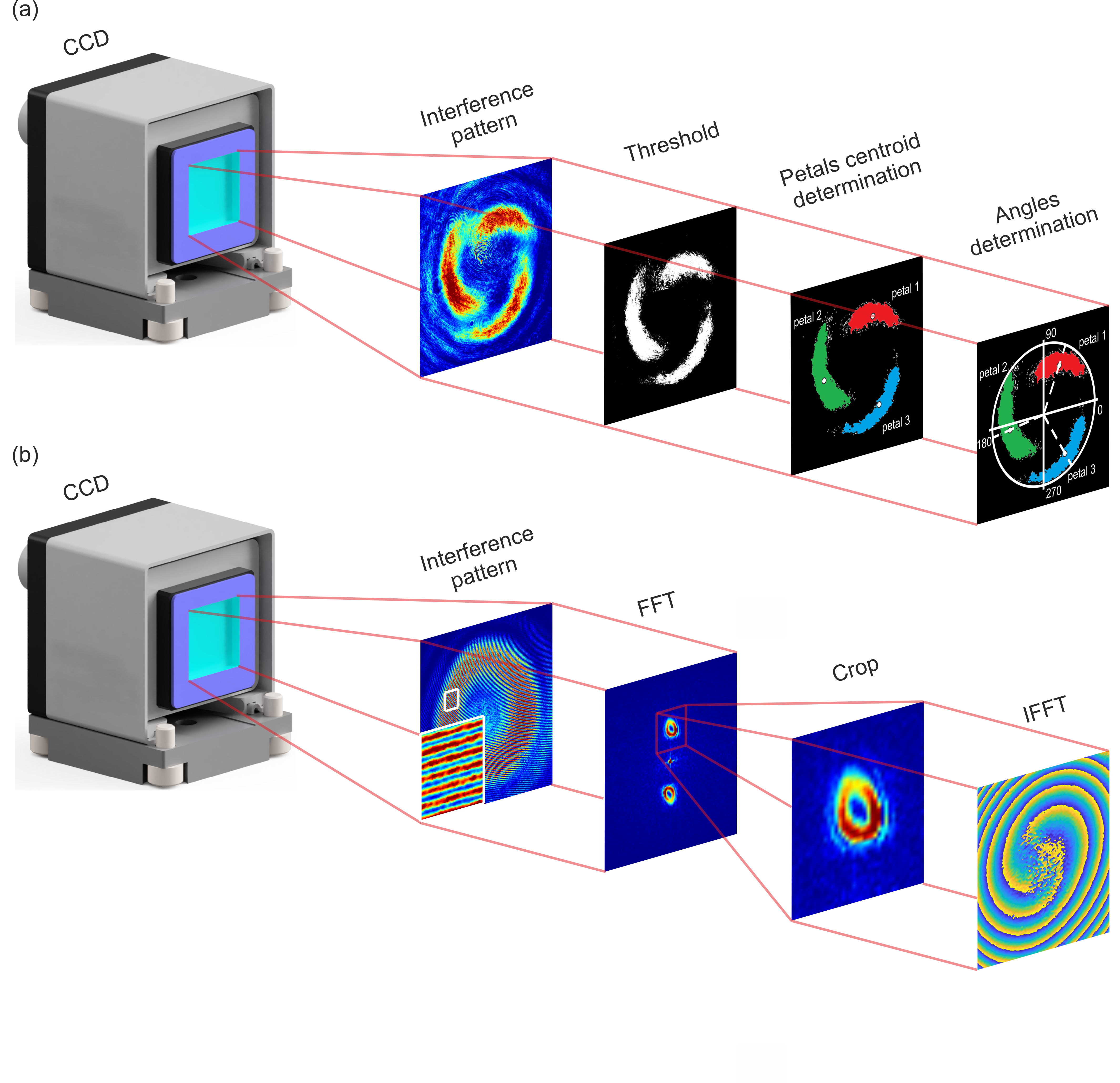}\\
    \caption{Schematic presentation of (a) Main steps and principles for determining the relative twist of petals in the on-axis regime; (b) Main steps and principles for the phase retrieval procedure in the off-axis regime.} 
      \label{fig:scheme}
\end{figure*}

\subsection*{APPENDIX E: Computational modelling}
Expressions (\ref{eq:LGbeamField})--(\ref{eq:LGbeamPhase}) describe intensity and phase distributions of the LG beam in the medium with $n=1$. To assess the conversion of OAM in the tissue-like scattering medium ($n\neq1$) a large set (e.g. $N_{ph} \sim 10^9$) of LG beam photon trajectories with starting points $\rho_{0_i},\phi_{0_i}, i\in[1 ... N_{ph}]$ is generated~\cite{Doronin2019}. Azimuths phase for each trajectory can be easily calculated as $-\ell\phi$ as follows from (\ref{eq:LGbeamPhase}). The length $L_i$ of each trajectory is estimated according to (\ref{eq:LGtrj}). Within cuvette (or within the sample) we estimate trajectory lengths as $\Delta L_i(r_{0_i},\varphi_{0_i})=L(r_{0_i},\varphi_{0_i},\zeta_2)-L(r_{0_i},\varphi_{0_i},\zeta_1)$, where $\zeta_2-\zeta_1$ corresponds to cuvette (sample) thickness in dimensionless units introduced in (\ref{eq:LGtrj}). Due to the non-zero contrast between refractive indices of free space, glass and interior of the cuvette (sample) the trajectories are additionally refracted according to Snell's law, slightly increasing $\Delta L_i$ values. 

We estimate cuvette (sample) influence on the LG beam propagation via phase retardation caused by the increasing path length of light within medium:
\begin{equation}
    \Delta\Psi_i = \dfrac{2\pi n \Delta L_i}{\lambda}.
        \label{eq:phaseEstimate}
\end{equation} 
Here, $\Delta\Psi_i$ corresponds to the phase retardation along $i$'th trajectory. By comparing the phase before cuvette $\Psi\vert_{L_i}$ and phase after cuvette $\Psi\vert_{L_i+\Delta L_i}=\Psi\vert_{L_i}+\Delta\Psi_i$ for different values of $n$, we obtain phase patterns that are twisted differently, as seen on Fig.~\ref{fig1}.

Upon computation of the intensity and phase beyond the cuvette (sample), it becomes feasible to derive the interference pattern of LG beams with a plane wave, yielding the characteristic petal pattern on the screen with $\lvert\ell\rvert$ petals. These petals undergo a twist in correspondence to variations in the refractive index $n$ within the cuvette interior. The quantification of this twist is elucidated in the ensuing procedure, and the obtained results exhibit notable concordance with experimental measurements, as illustrated in Fig.~\ref{fig1}.

The proposed computational methodology facilitates the assessment of pathlengths $\Delta L_i$ in scenarios where the cuvette (sample) interior manifests turbidity. In such instances, the trajectories of LG beams are construed in the context of Monte Carlo photons~\cite{Doronin2019}, undergoing multiple scattering events dictated by the medium's scattering coefficient $\mu_s$, absorption coefficient $\mu_a$, and anisotropy of scattering parameter $g$ ($g = \langle\cos\theta\rangle$).

\subsection*{APPENDIX F: Characterization of twist of the LG beam propagated through the medium}
The quantification of the relative OAM twist in the on-axis regime is established by defining the change in polar angle of a petal due to its rotation around the center of the LG beam. To ascertain this twist, a polar coordinate system is introduced, with the origin situated at the center of the LG beam for each recorded experimental image. Subsequently, a binary representation is generated for each image by binarizing the intensity values, with those exceeding a defined threshold set to one, and all other values set to zero. The luminance threshold is determined utilizing the Otsu approach~\cite{Otsu1979}, and can be manually adjusted either for an individual image or across a series of images to ensure clarity in visualizing each petal on the binary representation.

The identification of pixels corresponding to the outer boundaries of each illuminated region in the binary image is achieved through the utilization of the Moore-Neighbor tracing algorithm, adapted to adhere to Jacob's stopping criteria~\cite{Gonzalez2004}. The preeminent bright spots are associated with the interference pattern petals of the LG beam, contingent upon the selection of an optimal contrast threshold. All other bright spots are deemed artifacts, and their respective boundary pixels are systematically excluded from subsequent data analysis.

By discerning the boundary pixels, individual polygons are meticulously constructed for each petal, preserving their inherent geometric characteristics. For every polygon, the polar coordinates of its centroid are calculated. The derivation of final relative twist values is accomplished by subtracting the angular coordinates of the centroids between different images, considering the spatial orientation of the petals and ensuring a seamless $0$-to-$2\pi$ transition. A schematic representation delineating the underlying principles and key steps of this method is presented in Fig.~\ref{fig:scheme}-a.
\\
\subsection*{APPENDIX G: Phase retrieval}
In the off-axis regime, the retrieval of the LG beam phase is accomplished through a straightforward signal processing approach, as elucidated in the work by Vayalamkuzhi et al.~\cite{vayalamkuzhi2021transform}. The schematic illustrating the principal steps of this procedure is presented in Fig.~\ref{fig:scheme}-b. Initially, a fast Fourier transform is employed on the detected interference pattern with the primary objective of extracting the frequency spectrum. Subsequently, a judicious selection is made to isolate the pertinent frequency spectrum corresponding to the LG beam, demarcated by the red square in the FFT image.

\end{document}